# Tutorial: *a priori* estimation of sample size, effect size, and statistical power for cluster analysis, latent class analysis, and multivariate mixture models

Edwin S Dalmaijer [1]

**Affiliation**

1. School of Psychological Science, University of Bristol, United Kingdom

**Contact Details**

Dr Edwin S. Dalmaijer, School of Psychological Science, University of Bristol, 12a Priory Road, Bristol, BS8 1TU, United Kingdom. Email: edwin.dalmaijer@bristol.ac.uk

**Acknowledgements**

This tutorial took shape over the process of advising several colleagues with power analyses for clustering in pre-registrations and grant applications. Many thanks in particular to Dr Camilla Nord, Georgia Turner, Dr Amy Orben, and Prof Claire Haworth for their cluster power questions.

**Conflicts of interest**

The author declares that they have no competing interests (financial or other) that could have influenced or appeared to influence the work reported here.

# Abstract

Before embarking on data collection, researchers typically compute how many individual observations they should do. This is vital for doing studies with sufficient statistical power, and often a cornerstone in study pre-registrations and grant applications. For traditional statistical tests, one would typically determine an acceptable level of statistical power, (gu)estimate effect size, and then use both values to compute the required sample size. However, for analyses that identify subgroups, statistical power is harder to establish. Once sample size reaches a sufficient threshold, effect size is primarily determined by the number of measured features and the underlying subgroup separation. As a consequence, *a priory* computations of statistical power are notoriously complex. In this tutorial, I will provide a roadmap to determining sample size and effect size for analyses that identify subgroups. First, I introduce a procedure that allows researchers to formalise their expectations about effect sizes in their domain of choice, and use this to compute the minimally required number of measured variables. Next, I outline how to establish the minimum sample size in subgroup analyses. Finally, I use simulations to provide a reference table for the most popular subgroup analyses: k-means, Ward agglomerative hierarchical clustering, c-means fuzzy clustering, latent class analysis, latent profile analysis, and Gaussian mixture modelling. The table shows the minimum numbers of observations per expected subgroup (sample size) and features (measured variables) to achieve acceptable statistical power, and can be readily used in study design.

# Introduction

Many interesting research questions relate to identifying distinct subgroups within a larger dataset. For example, subgroup analyses can be used to describe areas of methane emissions (Malone et al., 2022), chemical concentrations in waters (Li et al., 2022), energy consumption in buildings (Díaz Redondo et al., 2020), jobs on high-performance computing systems (Halawa et al., 2020), glacier zones (Barzycka et al., 2023), Mediterranean dust events (Nissenbaum et al., 2023), microbial networks (Favila et al., 2022), and riverbed types (Kwon et al., 2023). In medical research, subgroup analyses have been used to identify risk factors for hospitalisation from clinical notes (Song et al., 2023), ways to provide informational and emotional patient support (Duimel et al., 2022); or to uncover clinically meaningful patient profiles in bipolar personality disorder (Wolf et al., 2023), compulsive behaviour (Den Ouden et al., 2022), low back pain (Wilson et al., 2023), polycystic ovary syndrome (Kiconco et al., 2023), or vasculitis (Lee et al., 2023). In cognitive and behavioural sciences, subgroup analyses are rapidly gaining in popularity compared to traditional statistical tests (Figure 1, using bibliobanana by Dalmaijer, Van Rheede, et al., 2021). They have been used to describe distinct types of dietary patterns (Hennessy et al., 2023), executive function difficulties (Bathelt et al., 2018), individuals with autism (Parlett-Pelleriti et al., 2022), intimate partner violence (Alexander & Johnson, 2023), resilience to socio-economic disadvantage (Dalmaijer, Gibbons, et al., 2021), responses to cognitive training (Rennie et al., 2019), search organisation (Benjamins et al., 2019), and self-harm (Uh et al., 2021).

      While the above all seems rather promising, subgroup analyses are not always appropriately applied. A major issue is that many studies fail to test against a one-group solution (Toffalini et al., 2022). In addition, subgroup analyses within the same domain frequently show inconsistent findings (Alexander & Johnson, 2023; Stein & Bomyea, 2023), and this is likely due to under-addressed methodological and statistical limitations.

      Some go so far as to suggest subgroup analysis should not be used in psychological research (Toffalini et al., 2022). Others outline the conditions in which subgroup analyses can be appropriate (Dalmaijer et al., 2020, 2022), and provide detailed guidelines on recommended workflows and common issues (Gao et al., 2023).

      Many of the highlighted problems come from researchers' lack of familiarity with the unintuitive statistical properties of subgroup analyses. Like traditional tests, their statistical power is determined by sample size and effect size (Dalmaijer et al., 2020, 2022). However, unlike traditional tests, effect size (and thus power) accumulates over measured variables (Dalmaijer et al., 2020, 2022). In theory, this complexity renders it difficult to compute statistical power for subgroup analyses, and one could even consider it inappropriate to even attempt a power analysis for such an exploratory tool (Gao et al., 2023). However, in practice, countless papers have reported findings from underpowered analyses that are likely to be false positives (Toffalini et al., 2022). This issue could have been avoided if researchers had the tools to estimate power *before* touching any data.

      In this tutorial, I outline a method to establish the required number of observations (sample size) and features (number of measured variables) *a priori* for popular subgroup analyses. These

include k-means clustering, agglomerative hierarchical clustering with Ward linkage, c-means fuzzy clustering, latent class analysis, latent profile analysis, and Gaussian mixture modelling. My goal is to encourage researchers to consider statistical power before they start new data collection or secondary data processing, and to give them the tools to determine whether subgroup analyses would be appropriate for their data.

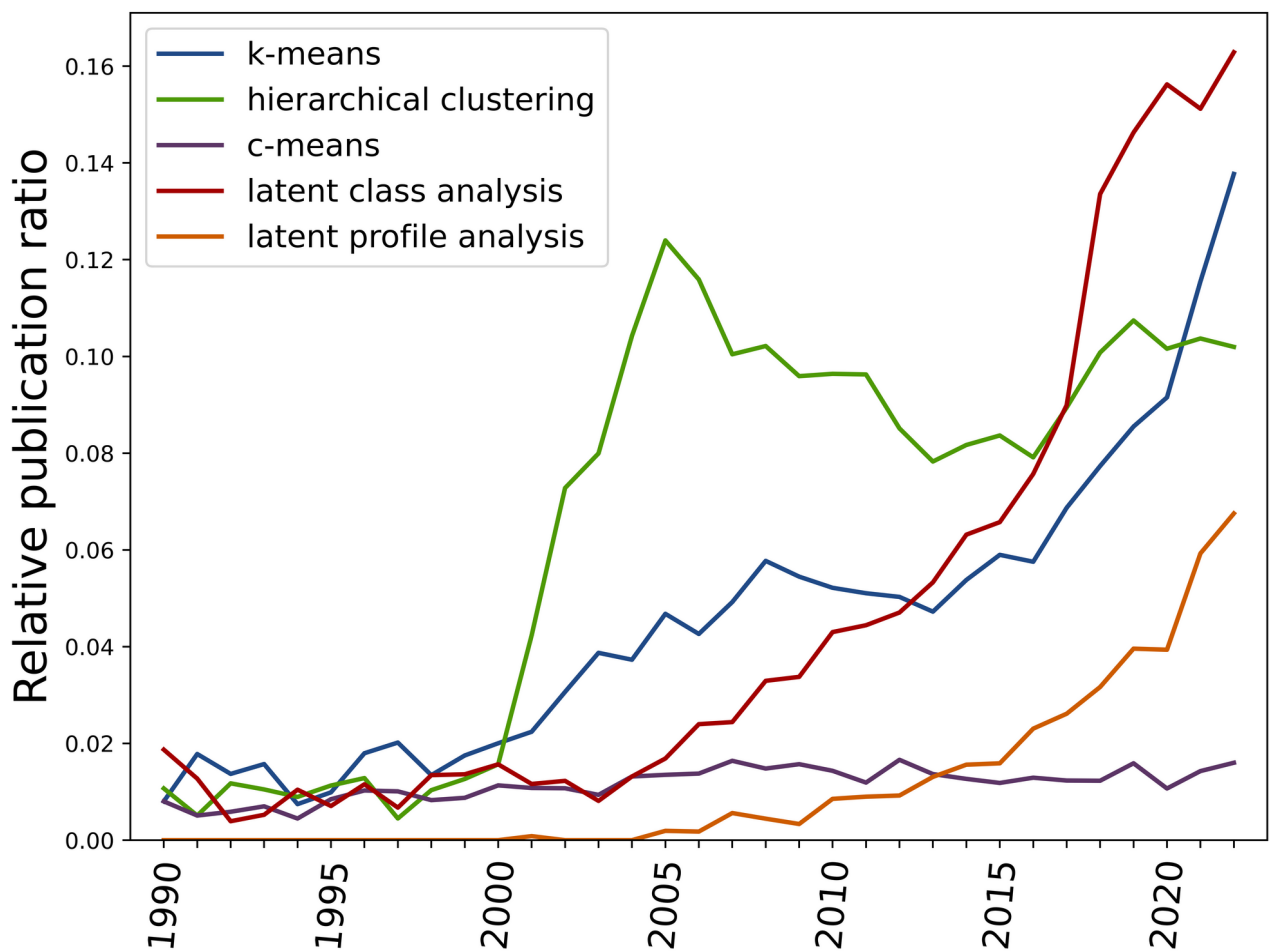

*Figure 1. The number of papers in journals indexed by PubMed that referenced a specific subgroup analysis, relative to the number of papers that referenced "t-test". This comparison shows that subgroup analyses have steadily gained in relative popularity (compared to traditional tests) since the year 2000, and an steeper increase since 2015. Plot generated using bibliobana (Dalmaijer, Van Rheede, et al., 2021).*

## What is statistical power in subgroup analyses?

In a frequentist framework, "power" refers to the probability that a test can detect a true positive. In simulations, it can be computed as the proportion of tests that result in a statistically significant result when a real effect was introduced. In subgroup analyses, power is a more elusive concept.

CLUSTER ANALYSIS SAMPLE SIZE AND POWER TUTORIAL

Where a traditional test offers a binary decision on there being an effect, a subgroup analysis offers much more information, including how many subgroups are present, and which observations belong to which subgroup. However, its central question can be reduced to a binary decision: Are any subgroups present in this data? The null hypothesis here is that all observations belong to a single group. Statistical power for subgroup analyses can thus be expressed as the probability of the single-group null hypothesis being correctly rejected. (Dalmaijer et al., 2020, 2022)

In simulations, statistical power can be estimated by generating two (or more) groups of data, employing a subgroup analysis, and then evaluating its outcome. Outcome evaluation entails computing the silhouette score (Rousseeuw, 1987) or its extension for probabilistic group membership (Campello & Hruschka, 2006), with scores over 0.5 rejecting the single-group null hypothesis. Power is then computed as the proportion of simulation runs that result in a silhouette score of 0.5 or higher.

**Alternatives to subgroup evaluation with silhouette scores?**

The silhouette score is not the only evaluation metric. However, it performs well compared to other validation indices across a wide variety of simulated and real datasets (Arbelaitz et al., 2013), even if it has some weaknesses for distributions with atypical shapes (Batool & Hennig, 2021).

Data in cognitive and behavioural research typically comprises partially overlapping normal distributions, and in this context evaluation with silhouette scores was without false positives in simulated single-group data, and with high accuracy in simulated multi-group data (Dalmaijer et al., 2020, 2022). Not all evaluation metrics are designed for this type of data; for example, the popular "gap statistic" is explicitly not intended to be used with overlapping groups (Tibshirani et al., 2001). This renders it, and methods that average across a variety of evaluation metrics (not all of which will be appropriate), unsuitable for our purposes.

Some subgroup analyses (e.g. mixture modelling) lend themselves to computing a goodness of fit (e.g. Bayesian Information Criterion) of a single-group null model, and comparing it with fits of multi-group models. While this could also support a binary rejection of the single-group null hypothesis, the beauty of this approach is that it offers a more continuous measure of evidence. This does not mean Bayesian evaluation is the most sensitive: In simulations with 1 to 4 subgroups (multivariate normal distributions with partial overlap), the thresholding of silhouette scores proved to be more accurate than the goodness-of-fit approach for detecting more than 2 subgroups with Gaussian mixture modelling without costing additional false positives (Dalmaijer et al., 2022).

In sum, the silhouette score has desirable properties in an evaluation metric. It can be computed for many (if not all) types of subgroup analyses, it works comparatively well in a wide variety of datasets, and its behaviour is well described for multivariate normal distributions with partial overlap (i.e. realistic data in cognitive and behavioural research).

# Estimating subgroup effect size

In a test of differences between two unrelated samples, effect size can be computed as the difference between the groups' means in standardised space (when the pooled standard deviation is equal to 1). Readers might recognise this as Cohen's *d*, and typical interpretations suggest that effects are very small at 0.01, small at 0.2, medium at 0.5, large at 0.8, very large at 1.2, and huge at 2.0 (Sawilowsky, 2009). Effect sizes for such differences will be referred to as *δ*.

The effect size in subgroup analyses can be defined as the distance between subgroup centroids. The higher this distance, the easier to identify each subgroup. When there are more than two subgroups, the lowest centroid distance determines how separable all subgroups are. Centroid distance is the effect size for subgroup analyses, and will be referred to as *Δ*.

The distance between centroids can be computed as the Euclidean distance between groups in standardised space (with feature variances of 1). This reflects an accumulation over differences between subgroups within each variable, and can be computed with Equation 1 (Dalmaijer et al., 2020, 2022).

$$(1) \quad \Delta = \sqrt{\sum_{i=1}^{p} \delta_i^2}$$

where *Δ* is the centroid distance (subgroup effect size), *p* is the number of features (measured variables), and $\delta_i$ is the difference effect size for feature *i*.

**Underlying (within-feature) effect size distributions**

While it is now possible to compute the subgroup effect size, the underlying effect sizes for each measured variable still need to be estimated. In a power analysis for a single difference, researchers typically turn to the published literature to estimate an effect size from previous studies or meta-analyses, where possible with a correction for publication (Van Aert & Van Assen, 2018) and outcome reporting bias (Van Aert & Wicherts, 2023). Ideally, one would do the same for each feature in a subgroup analysis, although it is unfeasible to do so for a large number of variables. Another obstacle is that it is typically unclear which subgroups are present in a population before a subgroup analysis is done, which makes it impossible to estimate any single underlying effect size.

While it may be impossible to estimate effect sizes within each measured variable, the distribution of effect sizes within a field of study can be established empirically: Szucs and Ioannidis (2017) extracted 26841 reported statistics from 3801 papers in cognitive neuroscience and psychology. Out of these, 17207 were "statistically significant" with an average effect size of *δ*=0.932, and 9634 were non-significant with an average *δ*=0.237, for a grand average of *δ*=0.683. The distribution of effect sizes was roughly exponential, with more papers finding smaller effects. The probability density of effect sizes can thus be approximated with Equation 2.

(2) $$f(\delta, \lambda) = \lambda e^{-\lambda \delta}$$

where $\delta$ is an effect size for a difference, and $\lambda$ is the exponential rate parameter.

The mean of an exponential distribution is given by the reciprocal of its $\lambda$ parameter (Equation 3). Assuming the underlying effect size distribution is indeed exponential, $\lambda$ can thus be estimated as the reciprocal of the grand mean from Szucs and Ioannidis (2017): 1 / 0.683 = 1.464 ≈ 1.5 (rounded).

(3) $$E[\delta] = \frac{1}{\lambda}$$

where $E[\delta]$ is the mean effect size, and $\lambda$ is the exponential rate parameter.

**Estimating centroid separation (subgroup effect size)**

The expected value for centroid separation (i.e. subgroup effect size) can be directly estimated by combining Equations 1 and 3. This offers researchers the opportunity to estimate subgroup effect size from the number of features (measured variables) included in the analysis and the anticipated $\lambda$ parameter for the underlying effect size distribution (Equation 4).

(4) $$\hat{\Delta} = \sqrt{p\left(\frac{1}{\lambda}\right)^2}$$

where $\hat{\Delta}$ *(hat)* is the estimated subgroup effect size, *p* is the number of features included in a subgroup analysis, and $\lambda$ is the exponential rate parameter.

**Choosing a λ value for subgroup effect size estimation**

The span of published effect sizes is substantial, and includes very large ($\delta$=1.2) to huge ($\delta$=2*)* effects (Szucs & Ioannidis, 2017). These are not *impossibly* large effects. For example, the difference in average height between men (M=174.91 cm, SD=2.51, n=5153) and women (M=162.03 cm, SD=2.37, n=5763) in the Finnish Twin Cohort Study amounts to an effect size of around $\delta$=5 (Silventoinen et al., 2000). However, they are *implausible* effect sizes due to the abundance of studies with small sample sizes in the literature (Szucs & Ioannidis, 2017). By definition, these only have power to detect large effects, and paired with file-drawers full of null results (Rosenthal, 1979) and publication bias (Smart, 1964), this has likely inflated reported effects sizes (Szucs & Ioannidis, 2017).

To account for this inflation, low effect sizes can be upweighted by increasing the magnitude of the *λ* parameter. Here, I used the particularly optimistic value of *λ*=0.75, which should only be used when many substantial effects are anticipated (e.g. data from entirely different animal or plant species). I also used *λ*=1.5 to match effect sizes published in psychological literature (Szucs & Ioannidis, 2017), *λ*=3 to only include up to very large effects, *λ*=6 to only include up to large effects, and *λ*=12 to only include up to medium effects (Figure 2).

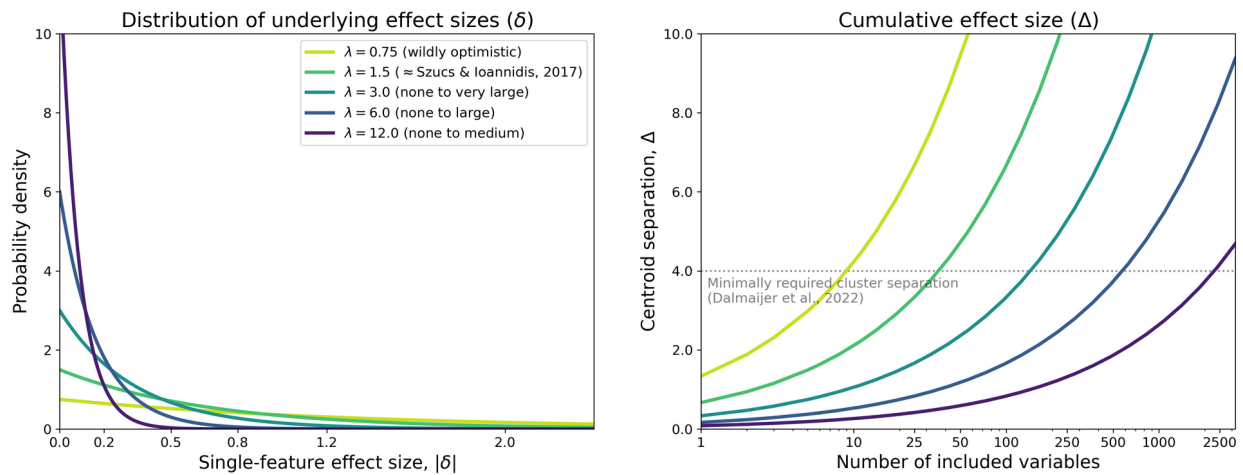

*Figure 2. Subgroup effect size as an accumulation of underlying effect sizes in different contexts. The left panel shows the probability densities (y-axis) as a function of effect sizes (difference between subgroups within a single feature) in five different contexts. The right panel shows estimated centroid separation (subgroup effect size) as an accumulation of single-feature effect sizes over features (x-axis, log-scaled). Effect size contexts ranged from highly liberal (lighter colours) to highly conservative (darker colours). Where they cross the dotted line reveals the number of features required to achieve an estimated subgroup effect size of Δ=4, which should typically offer sufficient statistical power to detect subgroups with k-means, c-means, HDBSCAN, or Gaussian mixture modelling (Dalmaijer et al., 2020, 2022). The main message of this figure is that many variables (500-2500) need to be included in context that are biased towards small effects.*

**Estimating the minimum number of variables**

With Equation 4, the centroid separation (subgroup effect size) can be estimated from the *λ* parameter of the underlying effect size distribution and the number of features included in an analysis. The *λ* parameter should be chosen on the basis of a datasets origin (see previous heading), but how to estimate the anticipated centroid separation or the required number of features is yet undiscussed. Choosing either parameter depends on whether data has already been collected.

If data has already been collected (e.g. when doing secondary data analysis), researchers can do a sensitivity analysis by using their chosen *λ* value and the number of features they can include (i.e. the number of useable variables in the dataset). Using Equation 4, this offers them an estimated value of subgroup effect size $\Delta$. By comparing this estimate with existing power computations for subgroup analyses published by Dalmaijer and colleagues (2020, 2022). The general recommendation here is that $\Delta>4$ should provide sufficient power to detect subgroups with k-means clustering, and $\Delta>3$ could be enough when using c-means or Gaussian mixture modelling are used (both assume multi-dimensional scaling is employed as a pre-processing step).

If data is yet to be collected, the number of variables to measure is undecided. The logic from the previous paragraph can be reversed: first a value of $\Delta$ for sufficient statistical power can be chosen, e.g. $\Delta=4$. Next, the minimum number of variables that should be included can be estimated by solving Equation 4 for the chosen $\Delta$, which results in Equation 5.

$$(5) \quad p = \frac{\Delta^2}{\left(\frac{1}{\lambda}\right)^2}$$

where $p$ is the number of features that should be included in a subgroup analysis, $\Delta$ is the required subgroup effect size, and $\lambda$ is the anticipated exponential rate parameter.

**Dimensionality reduction**

When performing a subgroup analysis on a dataset with many features, the "curse of dimensionality" can reduce performance even if a large distance between centroids exists. One way to combat this is by employing a dimensionality reduction algorithm to project high-dimensional data into a low-dimensional representation that can then be used in a subgroup analysis. An additional benefit of this approach is that multi-dimensional scaling can help increase centroid distances (Dalmaijer et al., 2020, 2022).

Dimensionality reduction algorithms for continuous data typically work by aiming to find a low-dimensionality representation in which relative distances between individual observations are analogous to those in the original data. These algorithms are unsuitable for categorical data, where the concept of "distance" does not apply in the same way. Dimensionality reduction can still be achieved to some extent by recoding categorical data into binary dummy variables (also know as "one hot encoding" in the machine-learning literature), which can then be reduced using "sketching" algorithms (Bera et al., 2023; Mitzenmacher et al., 2014).

Care should be taken in selecting a dimensionality reduction algorithm. For example, multi-dimensional scaling is computationally expensive and takes significant time on large datasets. Principal component analysis can be a faster alternative, but is more sensitive to covariance between features, particularly when underlying effect sizes are small (Figure 3). This can reduce centroid distances and thus hamper subgroup separability, especially in environments where within-variable differences are small and variables are likely to be correlated (e.g. in psychological and medical research).

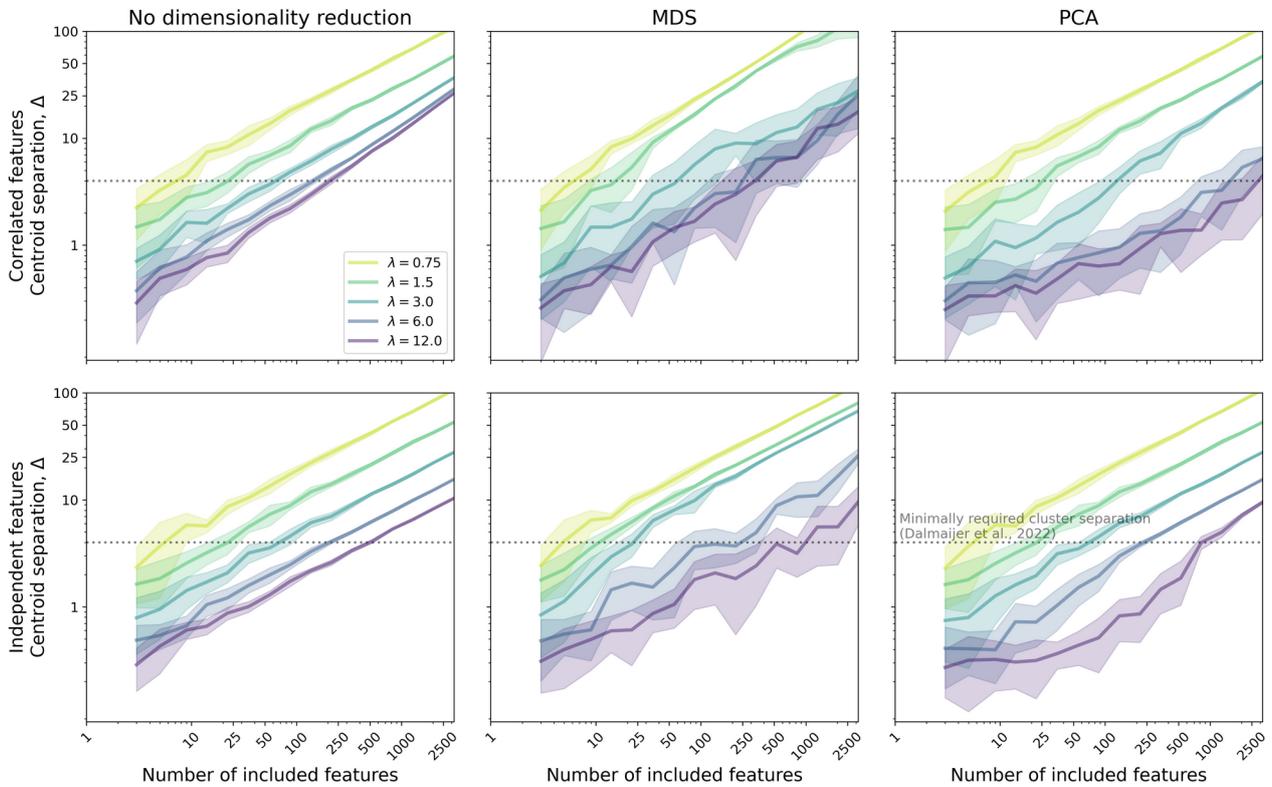

*Figure 3*. Distance (Δ, y-axis) between centroids computed before dimensionality reduction (left column), and after multi-dimensional scaling (MDS, middle column) or principal component analysis (PCA, right column) on correlated (top row) and independent (bottom row) features for different numbers of features (p, x-axis). Lighter colours indicate more optimistic effect size distributions (see Figure 2 for exact distributions). Solid lines are averages with shaded areas reflecting standard deviations across 10 simulations.

## Estimating sample size

Previous work suggests that *when centroid separation is sufficient*, the required sample size asymptotes at around *n*=30 per subgroup (Dalmaijer et al., 2020, 2022). Other recommendations suggest that the total sample size should be 70 times the number of features (Dolnicar et al., 2014); or even that the number of observations should be at least $2^p$, where *p* is the number of features (typically attributed to Formann, 1984). These recommendations vary substantially, and this will be addressed shortly.

How large the total sample should be depends on how many subgroups are hypothesised to exist, and what each subgroup's relative size is. For example, if three equally sized subgroups are hypothesised to exist, the total sample size would be 3*30 = 90. If two subgroups are hypothesised, but only 10% of the target population is expected to be a member of one of the subgroups, then at least 300 observations should be included (so that 10% of the sample results in *n*=30 for the smaller subgroup).

Further to the above, the preferred sample size depends on the number of included features. Two of the aforementioned recommendations are to increase the sample size as a function of the number of included features. According to these guidelines, inclusion of 100 features should require a sample size of 70*100 = 7000 or of $2^{100}$ = 1.27e30. These numbers are on entirely different orders of magnitude. The latter estimate is also impracticable: at the time of writing, the world population is estimated to be only 8e12 (United Nations, 2022), 1.58e17 times smaller than the intended sample size.

To address the issue of conflicting guidelines, I ran simulations that included *N*=30 to *N*=5000 per subgroup in various underlying effect size contexts. The results are covered in detail in the next section, but in short they suggest that while sample size should scale with the expected underlying effect size distribution, it can also be somewhat compensated by the number of features included in an analysis. This means that including more measured variables can actually reduce the sample size demands of a subgroup analysis.

## Reference table for popular analyses

To provide rough estimates for the required sample size and number of measured variables, I ran simulations with varying numbers of observations and features in different contexts of underlying effect sizes. Each simulated dataset was subjected to subgroup analyses k-means clustering, agglomerative hierarchical clustering with Ward linkage, c-means fuzzy clustering, latent profile analysis, and Gaussian mixture modelling. Categorical data was also simulated, and processed with latent class analysis. (Some of these analyses are mostly the same; see next section.)

Table 1 shows that the required numbers of observations and features increase as a function of the effect size context. Researchers who are planning a subgroup analysis in an environment with small underlying effect sizes should thus plan to employ a large sample size and to measure many variables. Where this is not possible, they should reconsider doing a subgroup analysis.

Note that the table offers rough estimates. Higher precision estimates would be derived from running simulations specific to planned pipelines, including preprocessing, dimensionality reduction, and the subgroup analysis of choice.

*Table 1*
This table offers rough estimates for the minimal sample size per expected group (n) and number of measured variables (p) for several popular subgroup analyses. The estimations were derived from simulations in five different distribution of underlying effect sizes (reflected by λ parameter). The "Published in psychology" column refers to the distribution of effect sizes observed in psychology-focussed journals (Szucs & Ioannidis, 2017). Columns are organised from most optimistic to most conservative, with λ reflecting the degree of weight towards null effects (see also Figure 2). Simulated data comprised uncorrelated variables that were projected into two dimensions using PCA (chosen for computational efficiency; not a preprocessing recommendation, see also Figure 3). Two options are given per row, one prioritising fewer features (top), and another prioritising fewer observations (bottom). Where the table states "No detection", subgroup analyses could not reliably identify the true two-subgroup solution over a single-group null model.

| | Wildly optimistic ($\lambda=0.75$) | Published in psychology ($\lambda=1.5$) | No to very large effects ($\lambda=3.0$) | No to large effects ($\lambda=6.0$) | No to medium effects ($\lambda=12$) |
|---|---|---|---|---|---|
| **K-means** | *n=75, p=9*  *n=30, p=14* | *n=30, p=36*  *n=30, p=56* | *n=100, p=144*  *n=30, p=225* | *n=750, p=324*  *n=200, p=576* | *n=2000, p=1296*  *n=500, p=2304* |
| **Ward** (agglomerative hierarchical clustering) | *n=75, p=9*  *n=30, p=14* | *n=30, p=36*  *n=30, p=56* | *n=100, p=144*  *n=30, p=225* | *n=200, p=576*  *n=100, p=900* | *n=5000, p=1296*  *n=500, p=2304* |
| **C-means** (fuzzy clustering) | *n=50, p=9*  *n=30, p=14* | *n=100, p=20*  *n=30, p=36* | *n=100, p=81*  *n=30, p=144* | *n=500, p=324*  *n=100, p=576* | *n=1000, p=1296*  *n=500, p=2304* |
| **Latent class analysis** | *n=50, p=9*  *n=30, p=14* | *n=150, p=20*  *n=100, p=36* | No detection | No detection | No detection |
| **Latent profile analysis** | *n=50, p=9*  *n=30, p=14* | *n=30, p=36*  *n=30, p=56* | *n=150, p=81*  *n=50, p=144* | *n=500, p=324*  *n=200, p=576* | *n=1000, p=1296*  *n=500, p=2304* |
| **Gaussian mixture modelling** | *n=75, p=9*  *n=30, p=14* | *n=30, p=36*  *n=30, p=56* | *n=50, p=144*  *n=30, p=225* | *n=500, p=324*  *n=200, p=576* | *n=1500, p=1296*  *n=500, p=2304* |

*Note*: the n values in this table are **per subgroup**, so the total sample size should be a multiple of the suggested values. For example, if three equally sized subgroups are expected, the planned sample size should be 3*n. If two subgroups are expected, with one being four times smaller than the other (i.e. an 80-20% divide of the population), the total sample size would be 4*n+n.

**Simulation details**

Simulated datasets comprised two equally sized subgroups, each represented by multivariate normal distributions. Effect sizes were drawn from exponential distributions with λ parameters of 0.75, 1.5, 3.0, 6.0, or 12.0 to match the distributions from Figures 2 and 3. For improved computational efficiency, features were uncorrelated and reduced through principal component analysis. These are not recommended assumptions or approaches, but chosen here only because they save on computing time by avoiding searches for positive-definite covariance matrices and dimensionality reduction through multi-dimensional scaling. For an idea of how feature correlations and dimensionality reduction algorithms impact subgroup analyses, see Figure 3 and Dalmaijer et al. (2022).

Feature numbers ($p$) were chosen for subgroup effect sizes $\Delta=3$, $\Delta=4$, or $\Delta=5$, using Equation 5. Sample sizes were run from $n=30$ to $n=5000$ per subgroup, resulting in $N=60$ to $N=10000$ for the full sample (comprising two subgroups). For each included combination of $n$ and $p$, 50 simulations were run. In each of these, one dataset was generated, and subjected to the following subgroup analyses.

The k-means clustering algorithm iteratively moves $k$ centroids until a stable solution is reached. In each iteration, observations are assigned to their closest centroid, and new centroids are defined as the average location of all their assigned observations (Lloyd, 1982).

In agglomerative hierarchical clustering, observations are joined as a function of their similarity. Similarity is typically defined by a distance metric (here Euclidean, but city block or cosine are also used), and compared against a linkage criterion. A commonly used criterion is Ward linkage, which minimises the increase in variance when observations are linked (Ward, 1963). The result is a hierarchy of solutions on the bottom of which each observation is its own cluster, and at the top of which all observations are part of the same cluster. Like for k-means, the accepted number of clusters is user-defined, or chosen as the solution with the highest silhouette score.

The c-means fuzzy clustering algorithm operates in much the same way as k-means, with one major difference: cluster membership is not binary. Instead, each observation is assigned a degree of belonging to each cluster (Bezdek, 1981; Dunn, 1973; Ross, 2010).

Latent class analysis is a latent mixture modelling approach in which predictors are binary (Bernoulli) or categorical (generalised Bernoulli or multinoulli) variables (Ferguson et al., 2020). Confusingly, the term "latent class analysis" has also been used to describe mixture modelling approaches more generally in the literature.

Latent profile analysis is a latent mixture modelling approach in which predictors are continuous variables that are assumed to be independent (Ferguson et al., 2020; Sterba, 2013). As with latent class analysis, the term "latent profile analysis" is also frequently used to describe mixture modelling more generally.

Gaussian mixture modelling is a latent mixture modelling approach in which predictors are continuous, and covariance structures can vary between subgroups. Here, all groups are allowed their own covariance structure. An alternative approach is to fit a single covariance matrix across all subgroups, which could help prevent over-fitting. Within mixture modelling approaches, the probability that an observation belongs to each cluster can be computed (analogous to fuzzy clustering).

For each subgroup analysis outside of latent class analysis, silhouette (Rousseeuw, 1987) or fuzzy silhouette (Campello & Hruschka, 2006) scores were computed. Power was then computed as the proportion of subgroup analysis that correctly rejected the single-group null hypothesis, i.e. the proportion of silhouette scores of 0.5 and over. Reported values for the required number of observations per subgroup and included features are for when analyses reached 90% power (45/50 simulations).

For latent class analysis, categorical values were simulated and then directly analysed via StepMix (Morin et al., 2023) without dimensionality reduction. This is because while it is possible to reduce categorical data, distances between observations are of a different quality and not always retained (Bera et al., 2023). Because silhouette scores operate on the same distance metric, their meaning is also hampered. In order to provide a fairer comparison, goodness of fit was computed for a null model (one group) and an alternative model (two subgroups). The two Bayesian Information Criteria were combined into a single Bayes Factor (Wagenmakers, 2007), for which values over 3 were taken as evidence for the two-subgroup model over the single-group null model.

Simulations were coded in Python version 3.8.10 (for a tutorial, see Dalmaijer, 2017), using NumPy version 1.22.3 (Harris et al., 2020), SciPy version 1.8.0 (Virtanen et al., 2020), scikit-learn version 1.1.0 (Pedregosa et al., 2011), scikit-fuzzy version 0.4.2, StepMix version 2.1.1 (Morin et al., 2023), the Intel extension for scikit-learn version 2023.2.1, and Matplotlib version 3.5.2 (Hunter, 2007). Code and generated data are freely available via GitHub: https://github.com/esdalmaijer/cluster_power_tutorial

## Conclusion

Before embarking on a subgroup analysis, researchers must choose how many observations (sample size) and features (measured variables) to include. The statistical power of subgroup analyses depends on subgroup effect size, which is quantified as the distance between subgroup centroids in standardised space. Without sufficient separation, subgroups cannot be detected. Centroid separation accumulates over measured variables, with small differences between subgroups within each variable determining centroid separation in multi-dimensional space. The effect sizes of these underlying differences can be modelled as an exponential distribution, with a $λ$ parameter estimated from the existing literature. Researchers can use this approach via the following workflow:

1. Choose the preferred subgroup analysis pipeline. This typically includes preprocessing, a dimensionality reduction algorithm, and a cluster analysis or mixture model.

2. Choose the subgroup effect size that should be minimally attained. For most approaches, this will be around $Δ=4$, but for some it might be as low as $Δ=3$ (e.g. c-means and mixture modelling).

3. Choose the expected underlying effect size distribution. This is an exponential distribution biased towards 0, with the degree of null bias determined by the $λ$ parameter. (Empirical estimates put at $λ$ at 1.5 in the psychological literature, but this is likely an under-estimation. A safer choice is $λ=3$ at a minimum.)

4. Estimate the required number of features using the chosen subgroup effect size ($\Delta$) and underlying effect size distribution parameter ($\lambda$) in Equation 5.

5. Simulate multivariate normal distributions with the computed number of variables, draw underlying effect sizes from an exponential distribution with the chosen $\lambda$ parameter to introduce a difference between two groups, subject simulated datasets to the chosen dimensionality reduction and subgroup analysis, and compute the solutions' silhouette scores. Increase the simulated number of participants until a high proportion (towards 1) of simulations results in a sufficient silhouette score (0.5 or over). The resulting number is the sample size for two equally sized subgroups; convert this to the total sample size using the number of subgroups hypothesised to exist in the data.

6. If the number of features is higher than the number of variables that can feasibly be collected (for new data) or that has already been collected (for secondary data), then subgroup analysis might not be a suitable approach.

Alternatively, researchers could use Table 1 for coarse estimates of the required number of observations per subgroup and the required number of features. This can be used as a first-pass on whether or not subgroup analysis is a feasible approach, i.e. whether the required sample size and number of measured variables are attainable in new data or present in existing data.

CLUSTER ANALYSIS SAMPLE SIZE AND POWER TUTORIALNissenbaum, D., Sarafian, R., Rudich, Y., & Raveh-Rubin, S. (2023). Six types of dust events in Eastern Mediterranean identified using unsupervised machine-learning classification. *Atmospheric Environment*, *309*, 119902. https://doi.org/10.1016/j.atmosenv.2023.119902

Parlett-Pelleriti, C. M., Stevens, E., Dixon, D., & Linstead, E. J. (2022). Applications of Unsupervised Machine Learning in Autism Spectrum Disorder Research: A Review. *Review Journal of Autism and Developmental Disorders*. https://doi.org/10.1007/s40489-021-00299-y

Pedregosa, F., Varoquaux, G., Gramfort, A., Michel, V., Thirion, B., Grisel, O., Blondel, M., Prettenhofer, P., Weiss, R., Dubourg, V., Vanderplas, J., Passos, A., Cournapeau, D., Brucher, M., Perrot, M., & Duchesnay, E. (2011). Scikit-learn: Machine learning in Python. *Journal of Machine Learning Research*, *12*, 2825–2830.

Rennie, J. P., Zhang, M., Hawkins, E., Bathelt, J., & Astle, D. E. (2019). Mapping differential responses to cognitive training using machine learning. *Developmental Science*. https://doi.org/10.1111/desc.12868

Rosenthal, R. (1979). The file drawer problem and tolerance for null results. *Psychological Bulletin*, *86*(3), 638–641. https://doi.org/10.1037/0033-2909.86.3.638

Ross, T. J. (2010). Chapter 10: Fuzzy Classification (subheading: Fuzzy c-Means Algorithm). In *Fuzzy logic with engineering applications* (3rd ed., pp. 352–353). Wiley.

Rousseeuw, P. (1987). Silhouettes: A graphical aid to the interpretation and validation of cluster analysis. *Journal of Computational and Applied Mathematics*, *20*, 53–65. https://doi.org/10.1016/0377-0427(87)90125-7

Sawilowsky, S. S. (2009). New Effect Size Rules of Thumb. *Journal of Modern Applied Statistical Methods*, *8*(2), 597–599. https://doi.org/10.22237/jmasm/1257035100

Silventoinen, K., Kaprio, J., Lahelma, E., & Koskenvuo, M. (2000). Relative effect of genetic and environmental factors on body height: Differences across birth cohorts among Finnish men and women. *American Journal of Public Health*, *90*(4), 627–630. https://doi.org/10.2105/AJPH.90.4.62720